\documentclass[reqno]{amsart}

\usepackage{booktabs}
\usepackage{IEEEtrantools}
\usepackage{latexsym}
\usepackage{graphicx}       
\usepackage{mathrsfs}
\usepackage{hyperref}
\usepackage{dsfont}

\usepackage[noadjust]{cite}
\usepackage{extarrows}

\usepackage{paralist}
\usepackage{capt-of}
\usepackage{xcolor}
\usepackage{diagbox}
\usepackage{cite}
\usepackage{subfigure}
\usepackage{amsmath,amssymb,amsfonts}
\usepackage{textcomp}
\usepackage{wasysym}
\usepackage{caption}

\usepackage{setspace}
\usepackage{cite}
\usepackage{color}

\begin{document}
	
\begin{abstract}
 Opacity has emerged as a central confidentiality notion for information-flow security in discrete event systems (DES), capturing the requirement that an external observer (intruder) should never be able to determine with certainty whether the system is, was, or will be in a secret state. This article provides a concise, newcomer-friendly overview of opacity in DES, emphasizing core definitions and the unifying estimation viewpoint behind major opacity notions,. We summarize representative verification techniques and highlight how different observation models reshape both the problem formulation and algorithmic structure. We then review principal enforcement paradigms, ranging from opacity-enforcing supervisory control to sensor activation/information release optimization and obfuscation/editing mechanisms. Beyond finite automata, we outline how opacity has been studied in richer models such as stochastic systems, timed systems, Petri nets, and continuous/hybrid dynamics, and we briefly survey applications in robotics, location privacy, and information services. Finally, we discuss selected open challenges, including solvability under incomparable information, scalable methods beyond worst-case complexity, and opacity under intelligent or data-driven adversaries.
\end{abstract}

\title[Opacity in Discrete Event Systems]{Opacity in Discrete Event Systems: A Perspective and Overview}

\author{Xiang Yin}

\address{School of Automation and Intelligent Sensing, Shanghai Jiao Tong University, Shanghai 200230.}
\email{yinxiang@sjtu.edu.cn} 
\maketitle

\section{Introduction}
In modern control systems, ensuring information security has become a central concern, especially in the broader context of information-flow security. As control systems grow increasingly interconnected and often span both cyber and physical components, the integrity and confidentiality of data flowing within and across these systems are critical to preventing unauthorized disclosure and malicious interference. Information-flow security aims to guarantee that sensitive information is not leaked to unauthorized entities through system behaviors or observable signals, thereby safeguarding reliability, safety, and performance.

Discrete event systems (DES) constitute one of the most fundamental formal models in control theory \cite{cassandras2007introduction}. They provide a structured abstraction for modeling and analyzing complex systems whose states evolve through discrete transitions triggered by events. DES models have been widely used across diverse application domains, including manufacturing systems, autonomous robotics, traffic management, and software-intensive systems, where high-level specifications and correctness requirements can be rigorously formulated and analyzed. This mathematical foundation makes DES particularly well suited for reasoning about control logic, correctness, and performance under partial observation and uncertainty.

Within this setting, the notion of \emph{opacity} has emerged as a prominent formalism for confidentiality in DES-based information-flow security. Informally, opacity captures the ability of a system to conceal secret information, typically modeled as secret states, secret behaviors, or secret predicates, from an external observer who can access some portion of the information flow, for example observable events, outputs, or logs. Over the past two decades, opacity has evolved into a rich family of security properties tailored to different disclosure risks and operational assumptions. Its integration into DES theory has enabled systematic analysis and design methodologies that aim to preserve confidentiality while maintaining desired functional behaviors.

This survey provides an introductory overview of opacity in DES, focusing on core concepts, theoretical foundations, representative problem formulations, and practical motivations. We review major opacity notions, such as current-state opacity and infinite-step opacity, along with common verification techniques for checking whether a DES satisfies a given opacity requirement. We also discuss principal enforcement paradigms, including supervisory control, dynamic observation mechanisms, and obfuscation or edit-based approaches, highlighting how these mechanisms trade off system utility and confidentiality. Finally, we outline selected challenges and open problems to clarify both the current maturity of the area and promising directions for future research, with particular attention to practical considerations that arise in deploying opacity-aware methods in emerging applications.

The author is fully aware that several comprehensive  survey papers on opacity in DES already exist  \cite{jacob2016overview,lafortune2018history,guo2020overview,joao2021analysis,hadjicostis2022cybersecurity,liu2022secure,yu2023survey,oliveira2023classification,yin2024formal,mangini2026recent}. The goal of this paper is not to compete with these works by exhaustively expanding the bibliography or providing a comprehensive technical account of every result. Rather, it is intended as a short, informal handbook for newcomers, aimed at quickly conveying the key ideas, typical methodologies, and the overall research landscape, so that readers can efficiently build initial intuition before moving on to more specialized treatments.

\section{Notions and Verifications of Opacity}

Since this is an overview paper, we will not formally define all the notations used, but will instead explain their meanings in the text. Readers are referred to \cite{cassandras2007introduction} for more technical details on discrete event systems.

\subsection{Basic Notions of Opacity}
We first review the basic formulations of opacity. 

\textbf{DES Model. }
A discrete event system (DES) is generally modeled as a tuple
\[
G = (X, \Sigma, \delta, X_0),
\]
where \( X \) represents the set of states, \( \Sigma \) is the set of events, \( \delta: X \times \Sigma \to 2^X \) is the transition function, and \( X_0 \subseteq X \) is the set of initial states.
In general, the state space \( X \) and the event space \( \Sigma \) can even be uncountably infinite. When both are finite, \( G \) is referred to as a finite state automaton.
A string is a finite sequence of events generated by \( G \), and we denote the language generated by \( G \) as \( L(G) \), which is the set of all possible strings it can generate.

\textbf{Information-Flow. }
In information-flow security analysis, it is assumed that the internal behavior \( s \in L(G) \) is not directly observable to an external observer, whom we refer to as a passive \emph{intruder} in this setting.
Formally, the information flow is modeled by a mapping
\[
P: \Sigma^* \to O^*
\]
that maps each internal string \( s \in L(G)\) to its corresponding information flow \( \alpha = P(s) \in O^* \), where \( O \) is the set of observation symbols.
A widely considered and simple case for \( P \) is the natural projection, where the observer can only see a set of observable events \( \Sigma_o \subseteq \Sigma \). Other types of observation mappings are discussed in Section~2.2.

\textbf{State Estimation. }
When the intruder observes an information flow \( \alpha \in P(L(G)) \), it can use this information, along with the system model \( G \), to infer the system's actual behavior. In particular, the set of all possible internal strings consistent with the observation is given by
$\mathcal{E}(\alpha) = P^{-1}(\alpha) \cap L(G)$.  
In many applications, the intruder is interested in estimating the system's state.
For finite state systems, state-based formulations are without loss of generality, as one can always convert language-based specifications into state-based ones, provided they are regular.
Using the estimated string \( \mathcal{E}(\alpha) \) and the system model \( G \), the intruder may be interested in past, present, or future behavior, leading to several types of state estimates. These are formally defined in \cite{yin2019estimation,hadjicostis2020estimation}, but here we describe them informally:
\begin{itemize}
\item 
\emph{Current State Estimate:} 
The set of all possible states the system can currently be in.\medskip
\item 
\emph{Initial State Estimate:} 
The set of all possible states the system could have started from.\medskip
\item 
\emph{Delayed State Estimate:} 
The set of all possible states the system could have been in \( K \) steps ago.\medskip
\item 
\emph{State Prediction:} 
The set of all possible states the system may reach \( K \) steps in the future.\medskip
\end{itemize}
Note that the number of steps in delayed state estimation or state prediction can be either counted by the number of observations or the actual events, depending on the specific problem setting. The initial state estimate is essentially a particular delayed state estimation, treating the delay as the entire observation length.

\textbf{Notions of Opacity. }
In opacity analysis, it is assumed that the system has certain secret behaviors, which can be captured by a sub-language \( L_S \subseteq L(G) \) or a subset of secret states \( X_S \subseteq X \).
As mentioned earlier, these two formulations are equivalent for regular specifications by refining the state space; thus, we will only consider the state-based formulation.
The general definition of opacity is that ``\textbf{The intruder can never know for sure that the system is, was, or will be in a secret state.}"
Depending on the aspect of the secret states the intruder is concerned with, several notions of opacity have been proposed in the literature.
Here, we provide generic definitions in text, though the formal definitions may vary in different contexts:
\begin{itemize}
\item 
\emph{Current-State Opacity (CSO):} 
Any reachable current state estimate cannot be a subset of the secret state, otherwise the intruder knows for sure that the system is in a secret state \cite{saboori2007notions,lin2011opacity}.\medskip
\item
\emph{Initial-State Opacity (ISO):} 
Any reachable initial state estimate cannot be a subset of the secret state, otherwise the intruder knows for sure that the system started from a secret state \cite{saboori2013verification,han2023strong}.\medskip
\item 
\emph{Initial-and-Final-State  Opacity (IFSO):} 
This generalizes the current and initial cases by requiring that the intruder can never determine specific initial-final state pairs \cite{wu2013comparative,masopust2025algorithms,miao2025always}.\medskip
\item 
\emph{\(K\)-Step Opacity (KSO):} 
For any observation, the delayed state estimation within the previous \( K \)-steps cannot be a subset of the secret state. This ensures that whenever the system passes through a secret state, the intruder cannot determine for sure that the system was in a secret state (either at a specific instant or along the trajectory) within the next \( K \) steps \cite{saboori2011verification2,falcone2015enforcement}.\medskip
\item 
\emph{Infinite-Step Opacity (InfSO):} 
When \( K \) in \( K\)-step opacity goes to infinity, it means that the intruder can never know for sure that the system is at a secret state, even in the long run \cite{saboori2011verification}.\medskip
\item 
\emph{Pre-Opacity (PO):} The intruder can never predict the visit to a secret state a certain number of steps ahead of time, before the system actually reaches the secret state \cite{yang2022secure}.
\end{itemize}
The relationship among notions of opacity and their transformations are also discussed in the literature \cite{lin2011opacity,wu2013comparative,balun2021comparing,wintenberg2022general}

\subsection{Verification of Opacity}

The opacity verification problem aims to determine, given a system model and an intruder (characterized by the information-flow mapping model), whether the system is opaque, i.e., whether there exists no secret-revealing string.
However, whether a string is secret-revealing depends on the existence of other non-secret strings having the same projection. Therefore, the opacity verification problem typically takes the following form:
\[
\forall s \in L_{\text{sec}}, \exists t \in L_{\text{non-sec}}: P(s) = P(t)
\]
or, equivalently, its negation:\[
\exists s \in L_{\text{sec}}, \forall t \in L_{\text{non-sec}}: P(s) \neq P(t).
\]
In terms of computational complexity, the opacity verification problem is generally PSPACE-hard, even for the standard natural projection setting, due to the quantifier alternation of the form \(\forall.\exists\).

To check the above conditions, the direct, though unavoidable, approach is to construct an observer-like structure over an exponential belief space or information-state space.
For instance, in the case of current-state opacity, we can build a current-state estimator over the state space \( 2^X \) and check for the existence of a secret-revealing state, which is a subset of secret states. This current-state estimator serves as a foundation and can be adapted to verify many other opacity notions with appropriate technical modifications such as state-space refinements.
For example, for initial-state opacity, the process becomes current-state opacity verification by augmenting the initial state into the state space. A more efficient approach using the reserved automaton technique is proposed in \cite{wu2013comparative}.
For pre-opacity, one can first perform state predictions in the original plant and then transform the problem into a current-state type problem. As shown in \cite{yin2017new}, the delayed state estimation problem is essentially a combination of current-state estimation and initial-state estimation. Therefore, \(K\)-step opacity and infinite-step opacity can be verified using a two-way observer structure—one based on the original plant and one based on the reversed plant. For other types of opacity, such as initial-final state opacity, some customized algorithms have been proposed, which are significantly more efficient \cite{masopust2025algorithms}. 
Recently, it has been shown in \cite{zhao2024unified} that some notions of opacity can also be verified efficient by using HyperLTL model checking techniques. 

\subsection{Different Information Structures in Opacity}
As noted above, our definition of the information-flow mapping \( P: \Sigma^* \to O^* \) is generic, and this specific form significantly influences the associated verification algorithm. Most of the standard algorithms mentioned earlier are based on the natural projection, which essentially represents a static observation mapping. However, the information mapping can be much more general. The following are some additional types of information mappings that have also been investigated in the literature.
\begin{itemize} 
    \item 
    \emph{Dynamic Observation.}  
    In dynamic observation models, the observability of an event is not predetermined but rather depends on the sequence of events leading up to the current point in time. This allows for the possibility that the observability of each event may change dynamically, influenced by the history of the system's behavior. In this context, an active information acquisition module, often referred to as the sensor activation policy, governs the decision on when to observe certain events.  Verification and synthesis of opacity under dynamic observations has been considered in \cite{cassez2012synthesis,yin2015codiagnosability,zhang2015maximum,behinaein2019optimal,yin2019synthesis,yin2019general}.\medskip
    \item 
    \emph{Orwellian  Observation.}  Orwellian observation introduces a more complex form of observability. In this model, the ability to observe an event depends not only on the past but also on future events, allowing the system to reinterpret past observations based on future outcomes. Some studies have proposed specific formulations, such as downgrading events, to make the problem decidable \cite{bryans2008opacity,mullins2014opacity,yeddes2016enforcing}. Recently,  a new model called the dynamic information release mechanism (DIRM) is proposed, which differs by allowing the state-dependent release of information, thus providing more flexibility in controlling the release of sensitive data \cite{hou2022framework}.\medskip  
    \item 
    \emph{Delayed Observation.} Unlike delayed state estimation, here the term ``delayed" refers to the observation itself. In networked settings, observations occur through channels, which may introduce delays or potentially cause them to arrive out of order. This can result in the observation sequence being either in correct order or misordered. The verification of opacity under delayed observation has also been studied in the literature \cite{lin2020information,yang2021opacity-net,zhang2021networked,wang2024distributed,wang2025enforcing}.
\end{itemize}
Note that some works also consider a nondeterministic observation or unreliable observation setting, in which each event may generate an observable symbol chosen from a possible set rather than being uniquely determined. This formalism can typically be reduced to the deterministic observation case by an appropriate state augmentation, for example by expanding the transition structure or enlarging the observation space \cite{takai2012verification,yin2016supervisor,yin2017initial,zhou2020detectability,takai2021general,li2024effect,dong2023uniform,zhang2025generalized}.

\textbf{Decentralized Opacity.}
The previous formulations primarily consider a single intruder who observes and estimates the behavior of the system. In some problems, however, the system is observed by multiple intruders who collaborate to infer the system's state. A typical scenario involves a decentralized decision architecture among the intruders, where the question arises as to what information each intruder can send to a fusion site and what rules the fusion site uses to determine the secrecy of the plant. Depending on the specific decentralized architecture, various notions of joint opacity have been explored in the literature \cite{paoli2012decentralized,tong2018decentralized,wu2018synthesis,ritsuka2025joint,sun2025construction}. These models account for the collaborative nature of the intruders and the interplay between their individual observations.

\textbf{Opacity in Modular Systems.}
In standard opacity analysis, the system  is typically considered as a monolithic plant. However, real-world large-scale systems are often naturally modular, composed of a set of local plants interconnected by event synchronizations. The opacity verification problem for such modular systems can, in general, be EXPSPACE-hard, revealing the inherent challenges associated with these structures \cite{yin2017verification,masopust2019complexity}. Nevertheless, some works have investigated specific features such as the decomposability of secrets, the compositionality of modules, and additional assumptions on private events, which have led to the development of more efficient modular verification algorithms \cite{mohajerani2019transforming,tong2019current,kalat2021modular,lennartson2022state,yang2022current}. These approaches leverage the modular nature of the system to simplify the opacity verification process, making it more tractable in practice.

\section{Enforcement of Opacity}
Given a system \( G \) and its associated information-flow mapping, the system may be verified to be non-opaque. To ensure information security, it is necessary to design enforcement mechanisms that enforce opacity. In general, there are two types of approaches to enforce opacity:
(i)
    One approach is to regulate the internal behavior of the system to avoid secret-revealing behavior under the original mapping.
  (ii) 
    The other approach is to modify the information mapping, either by regulating how information is acquired or released, or through artificial edits or encryption of the information available to the intruder.

\subsection{Supervisory Control of Opacity}
In the context of DES, the most fundamental approach to synthesizing a desired system is through supervisory control theory, initially proposed by Ramadge and Wonham \cite{ramadge1987supervisory}. In this framework, it is assumed that some events \( \Sigma_c \subseteq \Sigma \) are controllable, meaning their occurrence can be disabled by a supervisor, which has its own information-mapping \( M \), which is not necessarily identical to the intruder's mapping \( P \). The partial-observation supervisor is a function of the form \( S : M(L(G)) \to 2^{\Sigma_c} \) that dynamically disables/enables events online based on its observation \cite{lin1988observability,cieslak1988supervisory,yoo2006solvability,cai2014relative,yin2017synthesis,komenda2023hierarchical,yin2016synthesis,wonham2018supervisory,fokkink2024offline}, and we denote this system as \( L(S/G) \).

To correctly formulate the opacity supervisory control problem, one of the most essential aspects is whether or not the intruder knows the implementation of the supervisor. The presence or absence of this knowledge leads to fundamental differences in the problem. Specifically, when the system exhibits secret behavior \( s \in L(S/G) \cap L_{\text{sec}} \), if the intruder does not know the implementation of the supervisor, then it is permissible to provide a non-secret behavior \( t \in L(G) \cap L_{\text{non-sec}} \) from the open-loop system \( G \) such that \( P(s) = P(t) \). This ensures plausible deniability for the secret behavior, maintaining opacity. However, when the intruder knows the implementation of the supervisor, such non-secret behavior must be taken from the closed-loop system, i.e., \( t \in L(S/G) \cap L_{\text{non-sec}} \).
 
In the literature, the opacity supervisory control problem has been considered in various settings. For example, following the early works \cite{takai2008formula,dubreil2010supervisory,ben2011supervisory,saboori2011opacity}, 
most results consider the setting where the intruder knows the implementation of the supervisor \cite{darondeau2015enforcing,yin2015new,moulton2022using,zheng2025enforcement,dai2026optimal}. This setting faces a fundamental challenge as opacity is evaluated based on the intruder's state estimate, which is constructed based on the closed-loop system under control. However, the control law itself must still be synthesized. Such coupled control and estimation make the supervisory control problem open-ended, particularly when the observations of the supervisor and intruder are incomparable. Existing solvability results are often based on limited assumptions where their observations are either equal or one has strictly more information than the other \cite{dubreil2008opacity}. On the other hand, the unknown supervisory setting is much easier to handle, as one can couple control and estimation purely based on the open-loop system \cite{tong2018current,ma2021verification,xie2021secure}.

Regarding the various notions of opacity, their basic enforcement pipeline is similar, as opacity is essentially a safety problem concerning information \cite{yin2015uniform}. However, the key difference lies in the specific information-state form for each property. For instance, for current-state opacity, the standard power set space is sufficient. However, for $K$-step or infinite-step opacity, the information state is much more complex, as it requires tracking all possible delayed state estimate configurations encountered \cite{xie2020note,xie2023optimal}. Furthermore, for opacity involving prediction, such as pre-opacity, the information structure is more involved since the correctness depends on the future behavior of the closed-loop system, which has not yet been synthesized \cite{chen2023you,cui2025prediction}. It has also been shown that using non-deterministic supervisors can strictly increase the capacity to enforce opacity, but this results in a doubly exponential complexity in the information structure, as the supervisor needs to estimate the state estimate of the intruder \cite{xie2021opacity}.

Finally, in the supervisory control framework, since a feedback control channel is added, the information available to the intruder may also be reshaped. For example, the intruder can further observe part of the control decisions issued by the supervisor to improve its information regarding secret states. This leads to new opacity verification and control synthesis problems for opacity in networked control settings, which have also been investigated in the literature \cite{yin2019opacity,yang2021opacity,tai2025privacy}.

\subsection{Maximize Information under Opacity Constraint}
The supervisory control approach aims to regulate the internal behavior of the system under the original information-flow mapping. Alternatively, another approach to enforce opacity is to modify the observation mapping. In practice, the observation mapping captures the sensor activations that acquire information from the plant. For instance, the system may choose to dynamically turn sensors on and off for the purpose of energy efficiency \cite{wang2010minimization,sears2016minimal,sears2018computing,mao2024minimal}.
Clearly, the more sensors deployed, the more information the system gains, enabling it to perform more sophisticated tasks, such as distinguishing critical states. On the other hand, when too many sensors are activated, the likelihood of violating opacity increases. For example, sensors need to transmit the acquired information to the plant, and this information may be overheard by the intruder.
This issue is formulated as the dynamic sensor activation problem, which aims to synthesize an optimal dynamic sensor activation policy, either in terms of set inclusion or numerical metrics, such that opacity is satisfied \cite{cassez2012synthesis,yin2019synthesis,yin2019general,udupa2025synthesis}. The policy may also aim to meet other properties, such as distinguishability \cite{yin2019general,yin2017minimization}.

Another related problem is the maximal information release problem under opacity, where, upon each observation, the system decides whether or not to broadcast this event to the outside world. The goal is to release as much information as possible while ensuring the opacity of the system \cite{zhang2015maximum,behinaein2019optimal,liu2025distributed}. This problem is very similar to the sensor activation problem but with the following technical difference:
In the sensor activation problem, the decision module actively acquires information, and its dynamic observations are the same as those of the intruder. However, in the information release problem, the decision module passively observes information but actively determines what the intruder can access.
This difference makes the information release problem more challenging due to the incomparable information. In principle, we could consider a setting where the module both actively determines its own observations and selectively releases parts of them. However, this problem has not yet been considered in the literature, which presents an interesting research opportunity.

\subsection{Opacity Enforcement via Obfuscation }
The dynamic observation policy mentioned above essentially alters how actual information is obtained. In fact, the system can even manipulate the acquired information and release modified data to the intruder, which is referred to as obfuscation.
Along these lines, a basic approach is to use an insertion function, which inserts fictitious events between two consecutive actual events, making them indistinguishable from the intruder’s perspective \cite{wu2014synthesis,wu2015synthesis,ji2019enforcing,keroglou2020embedded,li2021extended,liu2022enforcement}. This insertion function has been extended in the literature in several directions. One such extension is the introduction of edit functions, where the mechanism can not only add additional events but also replace actual events with new fictitious ones \cite{liu2022opacity,li2024opacity,li2023opacity,duan2023edit,liu2025improved,duan2025obfuscation}.
Motivated by the advantages of non-deterministic supervisors, non-deterministic mechanisms have also been introduced to enhance the plausible deniability of the overall system \cite{ji2019opacity}. Additionally, similar to the known supervisory control setting, when the intruder knows the functionality of the edit functions, the original opacity enforcement may fail. This leads to the problem of synthesizing public-known obfuscation mechanisms that preserve opacity \cite{ji2018enforcement,liu2023opacity}.
Furthermore, the obfuscation mechanism should not only ensure opacity but also guarantee utility from the user's perspective \cite{wu2018synthesis,wintenberg2022dynamic,reis2026enforcing}. This sometimes leads to the design of an encode-decode mechanism between two users while ensuring opacity.
Finally, the use of obfuscation and supervisory control can be combined into a more powerful integrated framework when either approach alone cannot enforce opacity \cite{tai2023privacy,wintenberg2025integrating}.

\section{Opacity in Different State Models} 
Although finite state automata  provide a simple and fundamental model for the verification and analysis of opacity, this model fails to capture important aspects of more complex real-world dynamical systems. One key issue is its event-driven nature, which disregards timing information. Furthermore, FSA are typically used to model finite state spaces, whereas real-world systems can have countably infinite or even continuous state spaces. Additionally, such systems may evolve according to probabilistic distributions rather than purely logical rules. In the context of opacity, more complex system models have been considered to address these limitations. These models seek to capture the nuances of real-world applications, including stochastic, timed, continuous, and hybrid systems, each with their own challenges for ensuring opacity.

\subsection{Opacity in Stochastic Systems}
In stochastic systems, such as labeled Markov chains or probabilistic automata, opacity is defined differently depending on the probabilistic behavior of the system. Opacity in these systems is typically defined probabilistically, and it can be measured in various ways, depending on the specific characteristics of the system and the security requirements \cite{saboori2013current,berard2015quantifying,chen2016quantification,yin2017initial,keroglou2018probabilistic,berard2018opacity,yin2019infinite}.
There are two common formulations for defining opacity in stochastic systems. A simple approach is to ensure that the probability of revealing the secret (which is certain in a logical setting) is smaller than a given threshold. This opacity formulation can be addressed by combining a logical opacity verification algorithm, which constructs a suitable information structure to capture the state estimation, with probabilistic model checking techniques that compute the probability of executing certain language behaviors. This allows for the verification of opacity in a probabilistic context by estimating the likelihood of secret information being exposed.
Another, more challenging formulation requires that for any possible observation, the likelihood of being in a secret state must be smaller than a threshold. This formulation of opacity is more difficult to handle and often leads to undecidability results, as it is closely related to the probabilistic automaton emptiness problem. Essentially, this undecidability arises due to the infinite state-space in this setting. The belief state, representing the probabilistic distribution of states, forms an infinite space that cannot be exhaustively searched, making it computationally difficult to verify opacity. This has prompted the exploration of various approximations and restrictions to make opacity verification more tractable. Some techniques, such as belief state abstraction or sufficient conditions, have been proposed for verifying opacity in such settings, though these methods often involve trade-offs between precision and computational complexity \cite{wu2018privacy,ahmadi2018privacy,zheng2022privacy}.

In addition to verifying opacity, the controller synthesis problem, as well as other enforcement problems, have also been studied for stochastic systems such as Markov decision processes (MDPs) or probabilistic automata \cite{berard2015probabilistic,wu2018parameter,xie2021secure,cui2023security,udupa2025synthesis}. In many cases, this leads to multi-objective optimization problems, where a trade-off between opacity probability and other performance metrics, such as efficiency or control performance, needs to be considered.

\subsection{ Opacity in Timed Systems}
Timed systems introduce an additional layer of complexity, as they incorporate time as a factor influencing the system's state transitions. In these systems, the state does not only depend on the sequence of events but also on the timing of those events. This time-dependent behavior makes opacity more difficult to ensure, as the intruder may be able to infer information about the system based on the timing of its observations. For instance, in a timed automaton, a particular event might occur at different times in different executions of the system, potentially revealing information about the system's state. Even if the intruder cannot directly observe the states, the timing of events might still leak critical information.

For general timed automata, determining opacity is undecidable due to the unbounded nature of time \cite{cassez2009dark, wang2018opacity}. The complexity arises from the fact that time introduces infinite possibilities in how states evolve, making it computationally impossible to verify opacity for all configurations efficiently. 
However, various decidable variants of opacity in timed systems have been explored in the literature \cite{lai2021initial, ammar2021bounded, andre2022guaranteeing, andre2024bright, zhang2024state, dong2024verification, deng2025opacity, li2024verification, deng2025initial}. The decidability of opacity in timed systems largely depends on the types of time constraints imposed on the system. For example, timed automata with certain clock constraints may allow for a finite-state abstraction, making opacity verification possible within a finite context. By constraining the number of clocks or limiting the range of possible delays, one can reduce the complexity and render the opacity verification decidable.
Additionally, restrictive assumptions in discrete-time settings, bounded time horizons, or clock-reset mechanisms that are less expressive can also make the problem decidable. 
Beyond verification, the opacity enforcement problem has also been studied in the timed setting \cite{andre2022guaranteeing}.

\subsection{Opacity in Petri Nets}
Petri nets offer a powerful modeling tool for capturing infinite state spaces, making them well-suited for systems with concurrency and synchronization properties. Petri nets are particularly effective at representing systems where multiple components interact in parallel, as they provide a formal framework to model and analyze the flow of information and resources in such systems. Their structural properties allow for an efficient representation of state spaces, which is crucial when dealing with systems that have large or infinite state spaces.
However, the verification of opacity in Petri nets is generally undecidable due to the highly nonlinear nature of the reachability space, which cannot be efficiently represented or explored in its entirety \cite{bryans2005modelling, tong2017decidability, yin2017decidability, berard2018complexity, masopust2019deciding}.  Even for 1-safe Petri nets, which are the simplest class of bounded Petri nets with a finite state space, the reachable state space grows exponentially in the number of places.

Despite the undecidability and scalability issues, efficient analysis methods have been developed that do not require the full expansion of the state space. These methods leverage the structural properties of Petri nets to analyze systems efficiently. One of the key techniques is the bounded reachability graph (BRG) method, which allows for the analysis of system opacity by exploring a finite set of reachable states. The BRG method avoids the combinatorial explosion of state space by focusing only on a subset of reachable states that are relevant to the opacity verification process \cite{tong2016verification, ma2016basis, tong2022verification, ma2020marking, ma2021marking}. By using the BRG, it is possible to verify opacity without having to explicitly enumerate all possible states in the system. Additionally, linear programming methods have been used to verify opacity by modeling the state transitions and observations as linear constraints. These methods exploit the structure of Petri nets to efficiently check whether opacity conditions are met, even in relatively complex systems \cite{cong2018line, zhu2022online, basile2022necessary, basile2023assessment}.
In addition to opacity verification, the problem of opacity enforcement in Petri nets has also been addressed; see, e.g., \cite{basile2020noninterference,kohler2024enforcement, dong2025k}.

\subsection{Opacity in Continuous Dynamical Systems}
The system models considered above, although extending the state-space from finite to infinite, are still discrete as the space is countable. However, in cyber-physical systems, the state space is continuous, and their dynamics are typically described by differential equations. In these systems, opacity is closely related to the concept of unobservable subspaces in systems theory \cite{ramasubramanian2019notions, an2019opacity, john2024opacity}. 
Defining precise opacity in continuous dynamical systems is particularly challenging because the system's trajectories can be arbitrarily close to each other, yet correspond to different states. This issue is intrinsic to the continuous nature of the state space, where small variations in state variables can lead to behaviors that are indistinguishable from an observer with limited observational precision. In practice, the intruder may not have enough precision in their observations to distinguish between such closely related trajectories. Therefore, exact opacity is difficult to define because the observer cannot differentiate between trajectories that are nearly identical, leading to potential information leakage.
To address this, approximate opacity has been proposed. In this formulation, opacity is defined based on an approximation of the intruder's observations, where the system is considered opaque if the intruder cannot distinguish between behaviors of the system within a certain observational tolerance. This approach allows for a more practical way of defining opacity, as it accounts for the inherent limitations of observation precision \cite{yin2020approximate}.

For continuous systems with nonlinear dynamics, performing precise analysis of the system's behavior through analytical reachability analysis is generally infeasible.  Therefore, to make the problem tractable, the infinite state space must be abstracted in some manner.
A basic approach to manage this complexity is to define opacity-preserving simulation relations, which establish an equivalence between the opacity levels of two systems. These relations compare the observable behaviors of two systems and determine if one system can be substituted for another without violating opacity \cite{zhang2019opacity, hou2019abstraction, yin2020approximate, mizoguchi2021abstraction, hou2022abstraction, hou2023abstraction, qiao2025approximate}.
Recent research has extended this concept to stochastic systems, where the system's behavior is not deterministic but instead follows a probability distribution \cite{liu2020notion, liu2024approximate, qiao2025approximate}. Studies have also addressed large-scale systems composed of a set of local modules, further complicating the opacity verification process \cite{liu2021compositional, liu2021verification}.  

Despite the progress in these areas, challenges remain, particularly in terms of computational scalability and the accuracy of approximations. Traditional methods that rely on discretization can lead to significant computational overhead. More recent approaches, such as abstraction-free methods using barrier functions, have been proposed to overcome the limitations of discretization. These methods provide more efficient solutions for verifying opacity in continuous systems by directly using the system's dynamics and ensuring that the opacity conditions are satisfied without requiring an explicit discretization of the state space.
A key idea in these approaches is the use of an opacity barrier certificate as a sufficient condition for ensuring opacity. This certificate serves as a formal proof that the system will remain opaque under certain conditions. Such techniques have been applied not only to opacity verification \cite{liu2020verification, kalat2021modular} but also to control synthesis \cite{zhong2025secure}, ensuring that the system can be controlled in a way that maintains opacity throughout its operation. Additionally, these methods have been extended to a data-driven fashion, where opacity verification and control synthesis are achieved using data rather than a full model of the system \cite{murali2023data}. 

\section{Applications of Opacity and Related Topics}
In this section, we briefly discuss representative applications of opacity and several closely related topics that arise in information-flow security.

\subsection{Applications of Opacity}
Opacity has been applied broadly to information-flow security problems. Below, we briefly highlight a few representative application domains where opacity provides a natural confidentiality specification and motivates verification or enforcement methods.

\textbf{Autonomous Robots.} 
The notion of opacity, together with its associated synthesis algorithms, has been widely adopted in autonomous robot planning. Specifically, one typically assumes that an external observer  can only access partial observations of the robot’s executions, and the objective is to ensure that the robot’s behavior does not reveal whether certain secret states, regions, or task-relevant predicates have occurred, while simultaneously satisfying high-level specifications such as target navigation or more expressive linear temporal logic tasks \cite{saboori2011coverage,hadjicostis2018trajectory,wang2020hyperproperties,yang2020secure,zheng2023optimal,udupa2024planning,zhao2025no}. This line of research has also been extended to multi-robot systems, where each robot may possess its own private information that must be protected during coordinated planning and execution \cite{li2019coordinated,yu2022security,shi2023security}. Closely related security-aware planning problems have further been investigated in the context of combinatorial filter design, where the goal is to disclose sufficient information for safe and effective interaction with the environment while preserving individual privacy requirements that are conceptually aligned with opacity \cite{o2017concise,zhang2020you,phatak2023sensor}.

\textbf{Location Privacy.} 
Location privacy can be naturally modeled within the opacity framework by viewing a user’s true location/trajectory as the system state, and the data released to a location-based service (LBS) or smart environment (e.g., queried positions, sensor-triggered events, or time-labeled reports) as the adversary’s observation. The privacy requirement is then specified as a secret set of locations or predicates (such as ``the user is currently at home" or ``the user has visited a sensitive region"), and opacity demands that an external observer cannot determine with certainty—based on the released observations—whether the secret has occurred or holds at the current time. Under this formulation, opacity analysis provides a formal way to verify whether a given disclosure mechanism preserves location privacy, while opacity enforcement/synthesis enables the systematic design of controllers/filters (e.g., selective release, editing, or control-theoretic obfuscation) that balance service utility with privacy guarantees by ensuring the secret remains indistinguishable from non-secret behaviors \cite{wu2014ensuring,qin2023verification,danancher2015model,cerf2018control,goes2018demonstration}.

\textbf{Web Services and Cloud Computing.} 
In the context of web services and cloud computing, opacity can be applied to ensure data confidentiality and privacy by regulating what information is observable by external users, malicious parties, or service providers. In such environments, a user's actions, data requests, or resource utilization patterns may be tracked and potentially lead to the exposure of sensitive information, such as system configurations, user identities, or access behaviors. Opacity ensures that these secrets are not inferable from the observed interactions, even while maintaining the functionality of the system. This concept is especially relevant in cloud-based applications, where user data and processes are distributed across various platforms and accessed by different entities. By enforcing opacity, it is possible to prevent an adversary from deducing critical information about the underlying system state or the user’s sensitive data interactions, even if they can observe certain aspects of the service behavior. This has been explored in various studies focusing on the verification of opacity in cloud systems \cite{bourouis2016verification,zeng2019quantitative} and in securing industrial web applications deployed in cloud environments \cite{latha2024secure,mu2022verifying}.
 
\subsection{Opacity under Active Intruders} 
Most existing works on opacity focus on passive intruders, who simply observe the system without interacting with it. However, an intruder can be more powerful when it operates actively within the system \cite{yao2020attack,cui2025attack}. For example, an active intruder may override the control decisions of the system, altering its behavior and potentially violating opacity. Furthermore, similar to obfuscation techniques, an intruder may target the observation channel of the supervisory control system, modifying events or signals in such a way that the system reveals confidential information, thereby compromising its security \cite{su2018supervisor,meira2020synthesis}. Opacity under such active attack scenarios has been studied from both the supervisor's perspective, aiming to ensure opacity under all possible active attackers, and from the intruder's perspective, focusing on designing stealthy attack strategies. In the latter case, the intruder’s goal is often to remain undetected by the supervisor while subtly manipulating the system's behavior to mislead it into revealing sensitive information, thereby breaking the opacity guarantee \cite{tong2022polynomial,meira2023dealing,yao2024sensor,tai2025security}.

\subsection{Other Related Notions}
Opacity is closely related to a variety of observational properties studied in the DES literature, including diagnosability, detectability, observability, prognosability, and related notions. At a high level, these properties all concern what an external observer can infer, estimate, or predict about the internal evolution of a partially observed DES from its observation sequence. For instance, diagnosability requires that once a fault has occurred, its occurrence can be inferred within a finite delay from observations, whereas infinite-step opacity requires that once a secret has occurred, the observer can never determine with certainty that it occurred. In this sense, diagnosability can be viewed as an ``always eventually know" requirement, while infinite-step opacity corresponds to an ``always not know" requirement. Along similar lines, other information-flow security notions such as noninterference and its variants also formalize constraints on what can be inferred from observations, and their connections to opacity have been discussed in the literature \cite{ran2024noninterference,basile2020noninterference,hadj2005characterizing,yeddes2009modifying,ran2026non,basile2025verification}.
Many of these notions can also be composed to capture multi-level or igher-order inference patterns, for example requirements of the form ``the intruder should not know that it knows," which have been studied in the literature under the name \emph{higher-order opacity} \cite{cui2022you,cui2024better,miao2025protect}.
More recently, opacity and other observational properties have been revisited from a unified viewpoint through the lens of hyperproperties, which characterize system requirements in terms of relations among multiple executions rather than individual executions \cite{clarkson2010hyperproperties}. In the DES context, several classes of opacity have been expressed and analyzed using hyperproperty formalisms such as HyperLTL \cite{zhao2024unified}, enabling a more uniform specification language and facilitating comparisons across different security and inference-related properties.

\section{Conclusions}
Opacity has been studied extensively over the past two decades, spanning a wide range of opacity notions, system models, and enforcement mechanisms. Despite this rich body of work, the landscape is far from complete: certain combinations of opacity notions, modeling formalisms, and synthesis/verification objectives have not yet been systematically investigated. Moreover, the technical gaps are not uniform—some appear amenable to incremental extensions of existing frameworks, whereas others introduce fundamentally new challenges (e.g., changes in information structure, attacker capabilities, or computational complexity). In light of these open issues, the author believes that several promising and intellectually compelling directions remain for future research. Below, we highlight a few aspects that, from the author’s perspective, deserve particular attention:
\begin{itemize}
    \item 
    \textbf{Solvability with Incomparable Information.} 
    Within the current theoretical framework, opacity analysis and enforcement under a single observation mapping are relatively well understood. In contrast, many of the most challenging and long-standing open problems arise in settings with multiple observation sites that possess incomparable information. This includes standard supervisory control, dynamic information release, and decentralized variants of opacity. Even for the fundamental problem of opacity-enforcing supervisory control, decidability under general incomparable observation has remained open for nearly two decades. The precise decidability boundary and the associated complexity classes are still unclear, and this issue represents, in the author's view, one of the most fundamental theoretical challenges in the study of opacity. \medskip
    \item 
    \textbf{Beyond Worst-Case Complexity.}
    The computational complexity of opacity-related problems has been established for many system models, ranging from PSPACE-hardness for finite-state automata to undecidability results for Petri nets and timed systems. However, these results largely reflect worst-case behavior. The practical complexity of opacity verification and enforcement remains less well understood, especially in application-driven regimes. Progress in this direction calls for standardized benchmarks, broader numerical studies, scalable algorithms, and systematic comparisons that help characterize the gap between worst-case hardness and typical performance in practice.\medskip
    \item 
    \textbf{Opacity under Intelligent Attacks.}
    Standard formulations of opacity often assume a passive intruder who only observes system behavior. In open and interactive environments, however, an intruder may actively influence control decisions, manipulate observations, or strategically select actions to induce information leakage. While existing work has begun to address opacity under active intruders, it is often framed as worst-case analysis against an arbitrary attacker. In many practical scenarios, attackers are better modeled as strategic and resource-bounded agents with explicit utility functions, limited computational budgets, and specific goals. Developing a theory of opacity under intelligent attacks therefore requires richer models that capture rational adversarial behavior, potentially drawing on non-zero-sum game theory and incentive-aware security analysis. \medskip
    \item
    \textbf{Data-Cetric Opacity Analysis}.  
    Much of the current opacity theory is model-based, where both verification and inference are performed with respect to an explicit system model. In large-scale or highly complex systems, such models may be unavailable, inaccurate, or too costly to construct. In such cases, an intruder may rely primarily on operational data, combining offline datasets with online observations to infer secrets. This motivates a data-centric view of opacity that treats secrecy and inference through the lens of learning, estimation, and uncertainty. Advancing this direction likely requires new foundations that integrate system theory, information theory, and machine learning, including principled ways to quantify secrecy risks when only data-driven models are available.\medskip
    \item
    \textbf{Application to LLM Privacy.} 
    Finally, opacity may offer useful perspectives for  privacy in large language model (LLM) ecosystems. In many deployment settings, user prompts, interaction traces, or fine-tuning data may contain sensitive information, and model outputs can inadvertently expose private content through memorization or inference. It is of interest to explore whether opacity-inspired specifications can provide an operational and verifiable notion of confidentiality for such interactive AI systems, and whether enforcement mechanisms analogous to supervisory control or obfuscation can be adapted to mitigate leakage while preserving utility. Establishing suitable system abstractions, attacker models, and measurable secrecy objectives in this context constitutes an intriguing and potentially impactful direction.  
\end{itemize}

\bibliographystyle{elsarticle-num}   
\bibliography{ref} 

@article{cui2025attack,
  title={Attack-Resilient Supervisory Control of Discrete Event Systems Under Dynamic-Event-Protection Mechanisms},
  author={Cui, Bohan and Giua, Alessandro and Yin, Xiang},
  journal={IEEE Control Systems Letters},
  year={2025},
  publisher={IEEE}
}

@article{miao2025protect,
  title={Protect Your Knowledge: Epistemic Property Enforcement of Discrete Event Systems with Asymmetric Information},
  author={Miao, Shaowen and Cui, Bohan and Ji, Yiding and Yin, Xiang},
  journal={IEEE Control Systems Letters},
  year={2025},
  publisher={IEEE}
}

@article{cui2022you,
  title={You don't know what I know: On notion of high-order opacity in discrete-event systems},
  author={Cui, Bohan and Yin, Xiang and Li, Shaoyuan and Giua, Alessandro},
  journal={IFAC-PapersOnLine},
  volume={55},
  number={28},
  pages={135--141},
  year={2022},
  publisher={Elsevier}
}

@article{cui2024better,
  title={Better Late than Never: On Epistemic Diagnosability of Discrete Event Systems},
  author={Cui, Bohan and Ma, Ziyue and Giua, Alessandro and Yin, Xiang},
  journal={IFAC-PapersOnLine},
  volume={58},
  number={1},
  pages={174--179},
  year={2024},
  publisher={Elsevier}
}

@article{clarkson2010hyperproperties,
  title={Hyperproperties},
  author={Clarkson, Michael R and Schneider, Fred B},
  journal={Journal of Computer Security},
  volume={18},
  number={6},
  pages={1157--1210},
  year={2010},
  publisher={SAGE Publications Sage UK: London, England}
}

@article{ran2026non,
  title={Non-interference analysis of bounded labeled Petri nets},
  author={Ran, Ning and Wu, Zhengguang and Zhang, Shaokang and He, Zhou and Seatzu, Carla},
  journal={IEEE Transactions on Automatic Control},
  year={2026},
  publisher={IEEE}
}

@article{yeddes2009modifying,
  title={Modifying security policies for the satisfaction of intransitive non-interference},
  author={Yeddes, Moez and Lin, Feng and Hadj-Alouane, Nejib Ben},
  journal={IEEE Transactions on Automatic Control},
  volume={54},
  number={8},
  pages={1961--1966},
  year={2009},
  publisher={IEEE}
}

@article{hadj2005characterizing,
  title={Characterizing intransitive noninterference for 3-domain security policies with observability},
  author={Hadj-Alouane, Nejib Ben and Lafrance, St{\'e}phane and Lin, Feng and Mullins, John and Yeddes, Moez},
  journal={IEEE Transactions on Automatic Control},
  volume={50},
  number={6},
  pages={920--925},
  year={2005},
  publisher={IEEE}
}

@article{basile2025verification,
  title={Verification of K-step Non-Interference for Live Bounded and Reversible Discrete Event Systems Modeled with Petri Nets},
  author={Basile, Francesco and De Tommasi, Gianmaria and Dubbioso, Sara and Fiorenza, Federico},
  journal={IEEE Control Systems Letters},
  year={2025},
  publisher={IEEE}
}

@article{basile2020noninterference,
  title={Noninterference enforcement via supervisory control in bounded Petri nets},
  author={Basile, Francesco and De Tommasi, Gianmaria and Sterle, Claudio},
  journal={IEEE Transactions on Automatic Control},
  volume={66},
  number={8},
  pages={3653--3666},
  year={2020},
  publisher={IEEE}
}

@article{ran2024noninterference,
  title={Noninterference analysis of bounded Petri nets using basis reachability graph},
  author={Ran, Ning and Nie, Jingyao and Meng, Aiwen and Seatzu, Carla},
  journal={IEEE Transactions on Automatic Control},
  volume={69},
  number={10},
  pages={7159--7165},
  year={2024},
  publisher={IEEE}
}

@article{ji2019enforcing,
  title={Enforcing opacity by insertion functions under multiple energy constraints},
  author={Ji, Yiding and Yin, Xiang and Lafortune, St{\'e}phane},
  journal={Automatica},
  volume={108},
  pages={108476},
  year={2019},
  publisher={Elsevier}
}

@article{liu2022opacity,
  title={Opacity enforcement via attribute-based edit functions in the presence of an intended receiver},
  author={Liu, Rongjian and Lu, Jianquan and Hadjicostis, Christoforos N},
  journal={IEEE Transactions on Automatic Control},
  volume={68},
  number={9},
  pages={5646--5652},
  year={2022},
  publisher={IEEE}
}

@article{liu2025improved,
  title={Improved Synthesis of Nondeterministic Publicly-Known Edit Functions for Opacity Enforcement},
  author={Liu, Rongjian and Hadjicostis, Christoforos N},
  journal={IEEE Transactions on Automatic Control},
  year={2025},
  publisher={IEEE}
}

@article{ma2021verification,
  title={Verification and enforcement of strong infinite-and k-step opacity using state recognizers},
  author={Ma, Ziyue and Yin, Xiang and Li, Zhiwu},
  journal={Automatica},
  volume={133},
  pages={109838},
  year={2021},
  publisher={Elsevier}
}

@inproceedings{ben2011supervisory,
  title={Supervisory control for opacity of discrete event systems},
  author={Ben-Kalefa, Majed and Lin, Feng},
  booktitle={2011 49th Annual Allerton Conference on Communication, Control, and Computing (Allerton)},
  pages={1113--1119},
  year={2011},
  organization={IEEE}
}

@inproceedings{dubreil2008opacity,
  title={Opacity enforcing control synthesis},
  author={Dubreil, J{\'e}r{\'e}my and Darondeau, Philippe and Marchand, Herv{\'e}},
  booktitle={2008 9th international workshop on discrete event systems},
  pages={28--35},
  year={2008},
  organization={IEEE}
}

@article{fokkink2024offline,
  title={Offline supervisory control synthesis: taxonomy and recent developments},
  author={Fokkink, Wan and Goorden, Martijn},
  journal={Discrete Event Dynamic Systems},
  volume={34},
  number={4},
  pages={605--657},
  year={2024},
  publisher={Springer}
}

@article{lin1988observability,
  title={On observability of discrete-event systems},
  author={Lin, Feng and Wonham, Walter Murray},
  journal={Information sciences},
  volume={44},
  number={3},
  pages={173--198},
  year={1988},
  publisher={Elsevier}
}

@article{cieslak1988supervisory,
  title={Supervisory control of discrete-event processes with partial observations},
  author={Cieslak, Randy and Desclaux, C and Fawaz, Ayman S and Varaiya, Pravin},
  journal={IEEE Transactions on Automatic Control},
  volume={33},
  number={3},
  pages={249--260},
  year={1988},
  publisher={IEEE}
}

@article{liu2025distributed,
  title={Distributed Sensing and Information Transmission of Discrete-Event Systems With Edge Sensors},
  author={Liu, Yingying and Li, Shaoyuan and Hu, Jin and Yin, Xiang},
  journal={IEEE Transactions on Automatic Control},
  year={2025},
  publisher={IEEE}
}

@article{yin2017synthesis,
  title={Synthesis of maximally-permissive supervisors for the range control problem},
  author={Yin, Xiang and Lafortune, St{\'e}phane},
  journal={IEEE Transactions on Automatic Control},
  volume={62},
  number={8},
  pages={3914--3929},
  year={2017},
  publisher={IEEE}
}

@article{komenda2023hierarchical,
  title={Hierarchical supervisory control under partial observation: Normality},
  author={Komenda, Jan and Masopust, Tom{\'a}{\v{s}}},
  journal={IEEE Transactions on Automatic Control},
  volume={68},
  number={12},
  pages={7286--7298},
  year={2023},
  publisher={IEEE}
}

@article{yoo2006solvability,
  title={Solvability of centralized supervisory control under partial observation},
  author={Yoo, Tae-Sic and Lafortune, St{\'e}phane},
  journal={Discrete Event Dynamic Systems},
  volume={16},
  number={4},
  pages={527--553},
  year={2006},
  publisher={Springer}
}

@article{cai2014relative,
  title={Relative observability of discrete-event systems and its supremal sublanguages},
  author={Cai, Kai and Zhang, Renyuan and Wonham, Walter Murray},
  journal={IEEE Transactions on Automatic Control},
  volume={60},
  number={3},
  pages={659--670},
  year={2014},
  publisher={IEEE}
}

@article{yin2015codiagnosability,
  title={Codiagnosability and coobservability under dynamic observations: Transformation and verification},
  author={Yin, Xiang and Lafortune, St{\'e}phane},
  journal={Automatica},
  volume={61},
  pages={241--252},
  year={2015},
  publisher={Elsevier}
}

@article{wang2024distributed,
  title={Distributed fault diagnosis in discrete event systems with transmission delay impairments},
  author={Wang, Jiwei and Baldi, Simone and Yu, Wenwu and Yin, Xiang},
  journal={IEEE Transactions on Automatic Control},
  volume={69},
  number={8},
  pages={5508--5515},
  year={2024},
  publisher={IEEE}
}

@article{li2024effect,
  title={On the effect of dynamic event observations in distributed fault prognosis of discrete-event systems},
  author={Li, Bowen and Lu, Jianquan and Zhong, Jie and Wang, Yaqi},
  journal={IEEE Transactions on Automatic Control},
  volume={70},
  number={5},
  pages={2889--2901},
  year={2024},
  publisher={IEEE}
}

@article{takai2021general,
  title={A general framework for diagnosis of discrete event systems subject to sensor failures},
  author={Takai, Shigemasa},
  journal={Automatica},
  volume={129},
  pages={109669},
  year={2021},
  publisher={Elsevier}
}

@article{dong2023uniform,
  title={A uniform framework for diagnosis of discrete-event systems with unreliable sensors using linear temporal logic},
  author={Dong, Weijie and Yin, Xiang and Li, Shaoyuan},
  journal={IEEE Transactions on Automatic Control},
  volume={69},
  number={1},
  pages={145--160},
  year={2023},
  publisher={IEEE}
}

@article{zhou2020detectability,
  title={Detectability of discrete-event systems under nondeterministic observations},
  author={Zhou, Lei and Shu, Shaolong and Lin, Feng},
  journal={IEEE Transactions on Automation Science and Engineering},
  volume={18},
  number={3},
  pages={1315--1327},
  year={2020},
  publisher={IEEE}
}

@article{zhang2025generalized,
  title={Generalized Critical Observability of Labeled Petri Nets Under Nondeterministic Observations},
  author={Zhang, Yuling and Hadjicostis, Christoforos N and Li, Zhiwu},
  journal={IEEE Transactions on Automation Science and Engineering},
  year={2025},
  publisher={IEEE}
}

@article{yin2016supervisor,
  title={Supervisor synthesis for mealy automata with output functions: A model transformation approach},
  author={Yin, Xiang},
  journal={IEEE Transactions on Automatic Control},
  volume={62},
  number={5},
  pages={2576--2581},
  year={2016},
  publisher={IEEE}
}

@article{takai2012verification,
  title={Verification of codiagnosability for discrete event systems modeled by Mealy automata with nondeterministic output functions},
  author={Takai, Shigemasa and Ushio, Toshimitsu},
  journal={IEEE Transactions on Automatic Control},
  volume={57},
  number={3},
  pages={798--804},
  year={2012},
  publisher={IEEE}
}

@book{cassandras2007introduction,
  title={Introduction to discrete event systems},
  author={Cassandras, Christos G and Lafortune, Stephane},
  year={2007},
  publisher={Springer}
}

@article{yin2019estimation,
  title={Estimation and Verification of Partially Observed Discrete-Event Systems},
  author={Yin, Xiang},
  journal={Wiley Encyclopedia of Electrical and Electronics Engineering},
  pages={1--12},
  year={2019},
  publisher={Wiley Online Library}
}

@inproceedings{saboori2007notions,
  title={Notions of security and opacity in discrete event systems},
  author={Saboori, Anooshiravan and Hadjicostis, Christoforos N},
  booktitle={2007 46th IEEE Conference on Decision and Control},
  pages={5056--5061},
  year={2007},
  organization={IEEE}
}

@article{lin2011opacity,
  title={Opacity of discrete event systems and its applications},
  author={Lin, Feng},
  journal={Automatica},
  volume={47},
  number={3},
  pages={496--503},
  year={2011},
  publisher={Elsevier}
}

@article{saboori2013verification,
  title={Verification of initial-state opacity in security applications of discrete event systems},
  author={Saboori, Anooshiravan and Hadjicostis, Christoforos N},
  journal={Information Sciences},
  volume={246},
  pages={115--132},
  year={2013},
  publisher={Elsevier}
}

@article{masopust2025algorithms,
  title={On algorithms verifying initial-and-final-state opacity: Complexity, special cases, and comparison},
  author={Masopust, Tom{\'a}{\v{s}} and Osi{\v{c}}ka, Petr},
  journal={Automatica},
  volume={174},
  pages={112171},
  year={2025},
  publisher={Elsevier}
}

@article{falcone2015enforcement,
  title={Enforcement and validation (at runtime) of various notions of opacity},
  author={Falcone, Ylies and Marchand, Herv{\'e}},
  journal={Discrete Event Dynamic Systems},
  volume={25},
  number={4},
  pages={531--570},
  year={2015},
  publisher={Springer}
}

@article{zhao2024unified,
  title={A unified framework for verification of observational properties for partially-observed discrete-event systems},
  author={Zhao, Jianing and Li, Shaoyuan and Yin, Xiang},
  journal={IEEE Transactions on Automatic Control},
  volume={69},
  number={7},
  pages={4710--4717},
  year={2024},
  publisher={IEEE}
}

@article{cassez2012synthesis,
  title={Synthesis of opaque systems with static and dynamic masks},
  author={Cassez, Franck and Dubreil, J{\'e}r{\'e}my and Marchand, Herv{\'e}},
  journal={Formal Methods in System Design},
  volume={40},
  number={1},
  pages={88--115},
  year={2012},
  publisher={Springer}
}

@article{yin2019synthesis,
  title={Synthesis of dynamic masks for infinite-step opacity},
  author={Yin, Xiang and Li, Shaoyuan},
  journal={IEEE Transactions on Automatic Control},
  volume={65},
  number={4},
  pages={1429--1441},
  year={2019},
  publisher={IEEE}
}

@inproceedings{yeddes2016enforcing,
  title={Enforcing opacity with Orwellian observation},
  author={Yeddes, Moez},
  booktitle={2016 13th international workshop on discrete event systems (WODES)},
  pages={306--312},
  year={2016},
  organization={IEEE}
}

@article{lin2020information,
  title={Information control in networked discrete event systems and its application to battery management systems},
  author={Lin, Feng and Wang, Le Yi and Chen, Wen and Wang, Weilin and Wang, Fei},
  journal={Discrete Event Dynamic Systems},
  volume={30},
  number={2},
  pages={243--268},
  year={2020},
  publisher={Springer}
}

@article{zhang2021networked,
  title={Networked opacity for finite state machine with bounded communication delays},
  author={Zhang, Zhipeng and Shu, Shaolong and Xia, Chengyi},
  journal={Information Sciences},
  volume={572},
  pages={57--66},
  year={2021},
  publisher={Elsevier}
}

@inproceedings{tong2018decentralized,
  title={Decentralized opacity enforcement in discrete event systems using supervisory control},
  author={Tong, Yin and Cai, Kai and Giua, Alessandro},
  booktitle={2018 57th Annual Conference of the Society of Instrument and Control Engineers of Japan (SICE)},
  pages={1053--1058},
  year={2018},
  organization={IEEE}
}

@article{ritsuka2025joint,
  title={Joint opacity and opacity against state-estimate-intersection-based intrusion of discrete-event systems},
  author={Ritsuka, K and Lafortune, St{\'e}phane and Lin, Feng},
  journal={Automatica},
  volume={176},
  pages={112136},
  year={2025},
  publisher={Elsevier}
}

@article{sun2025construction,
  title={Construction of Decentralized State Estimator Based on Partially Ordered Observation Sequences},
  author={Sun, Dajiang and Hadjicostis, Christoforos N and Li, Zhiwu},
  journal={IEEE Transactions on Automatic Control},
  year={2025},
  publisher={IEEE}
}

@inproceedings{wu2018synthesis,
  title={Synthesis of insertion functions to enforce decentralized and joint opacity properties of discrete-event systems},
  author={Wu, Bo and Dai, Jin and Lin, Hai},
  booktitle={2018 Annual American Control Conference (ACC)},
  pages={3026--3031},
  year={2018},
  organization={IEEE}
}

@inproceedings{paoli2012decentralized,
  title={Decentralized opacity of discrete event systems},
  author={Paoli, Andrea and Lin, Feng},
  booktitle={2012 American Control Conference (ACC)},
  pages={6083--6088},
  year={2012},
  organization={IEEE}
}

@article{yang2021opacity-net,
  title={Opacity of networked discrete event systems},
  author={Yang, Jingkai and Deng, Weilin and Qiu, Daowen and Jiang, Cheng},
  journal={Information Sciences},
  volume={543},
  pages={328--344},
  year={2021},
  publisher={Elsevier}
}

@article{bryans2008opacity,
  title={Opacity generalised to transition systems},
  author={Bryans, Jeremy W and Koutny, Maciej and Mazar{\'e}, Laurent and Ryan, Peter YA},
  journal={International Journal of Information Security},
  volume={7},
  number={6},
  pages={421--435},
  year={2008},
  publisher={Springer}
}

@article{joao2021analysis,
  title={Analysis and Control for Resilience of Discrete Event Systems},
  author={Jo{\~a}o, Carlos Basilio and Christoforos, N Hadjicostis and Rong, Su},
  journal={Foundations and Trends{\textregistered} in Systems and Control},
  volume={8},
  number={4},
  pages={285--443},
  year={2021},
  publisher={Emerald Publishing Limited}
}

@inproceedings{hadjicostis2022cybersecurity,
  title={Cybersecurity and supervisory control: A tutorial on robust state estimation, attack synthesis, and resilient control},
  author={Hadjicostis, Christoforos N and Lafortune, St{\'e}phane and Lin, Feng and Su, Rong},
  booktitle={2022 IEEE 61st Conference on Decision and Control (CDC)},
  pages={3020--3040},
  year={2022},
  organization={IEEE}
}

@article{mullins2014opacity,
  title={Opacity with orwellian observers and intransitive non-interference},
  author={Mullins, John and Yeddes, Moez},
  journal={IFAC Proceedings Volumes},
  volume={47},
  number={2},
  pages={344--349},
  year={2014},
  publisher={Elsevier}
}

@article{hou2022framework,
  title={A framework for current-state opacity under dynamic information release mechanism},
  author={Hou, Junyao and Yin, Xiang and Li, Shaoyuan},
  journal={Automatica},
  volume={140},
  pages={110238},
  year={2022},
  publisher={Elsevier}
}

@article{zhang2015maximum,
  title={Maximum information release while ensuring opacity in discrete event systems},
  author={Zhang, Bo and Shu, Shaolong and Lin, Feng},
  journal={IEEE Transactions on Automation Science and Engineering},
  volume={12},
  number={3},
  pages={1067--1079},
  year={2015},
  publisher={IEEE}
}

@article{keroglou2018probabilistic,
  title={Probabilistic system opacity in discrete event systems},
  author={Keroglou, Christoforos and Hadjicostis, Christoforos N},
  journal={Discrete Event Dynamic Systems},
  volume={28},
  number={2},
  pages={289--314},
  year={2018},
  publisher={Springer}
}

@article{yin2019infinite,
  title={Infinite-step opacity and K-step opacity of stochastic discrete-event systems},
  author={Yin, Xiang and Li, Zhaojian and Wang, Weilin and Li, Shaoyuan},
  journal={Automatica},
  volume={99},
  pages={266--274},
  year={2019},
  publisher={Elsevier}
}

@article{berard2015probabilistic,
  title={Probabilistic opacity for Markov decision processes},
  author={B{\'e}rard, B{\'e}atrice and Chatterjee, Krishnendu and Sznajder, Nathalie},
  journal={Information Processing Letters},
  volume={115},
  number={1},
  pages={52--59},
  year={2015},
  publisher={Elsevier}
}

@article{wu2018privacy,
  title={Privacy verification and enforcement via belief abstraction},
  author={Wu, Bo and Lin, Hai},
  journal={IEEE Control Systems letters},
  volume={2},
  number={4},
  pages={815--820},
  year={2018},
  publisher={IEEE}
}

@article{zheng2022privacy,
  title={Privacy-preserving pomdp planning via belief manipulation},
  author={Zheng, Wei and Jung, Taeho and Lin, Hai},
  journal={IEEE Control Systems Letters},
  volume={6},
  pages={3415--3420},
  year={2022},
  publisher={IEEE}
}

@inproceedings{ahmadi2018privacy,
  title={Privacy verification in POMDPs via barrier certificates},
  author={Ahmadi, Mohamadreza and Wu, Bo and Lin, Hai and Topcu, Ufuk},
  booktitle={2018 IEEE Conference on Decision and Control (CDC)},
  pages={5610--5615},
  year={2018},
  organization={IEEE}
}

@article{yin2017initial,
  title={Initial-state detectability of stochastic discrete-event systems with probabilistic sensor failures},
  author={Yin, Xiang},
  journal={Automatica},
  volume={80},
  pages={127--134},
  year={2017},
  publisher={Elsevier}
}

@article{saboori2013current,
  title={Current-state opacity formulations in probabilistic finite automata},
  author={Saboori, Anooshiravan and Hadjicostis, Christoforos N},
  journal={IEEE Transactions on automatic control},
  volume={59},
  number={1},
  pages={120--133},
  year={2013},
  publisher={IEEE}
}

@book{hadjicostis2020estimation,
  title={Estimation and inference in discrete event systems},
  author={Hadjicostis, Christoforos N},
  year={2020},
  publisher={Springer}
}

@article{yin2017minimization,
  title={Minimization of sensor activation in decentralized discrete-event systems},
  author={Yin, Xiang and Lafortune, St{\'e}phane},
  journal={IEEE Transactions on Automatic Control},
  volume={63},
  number={11},
  pages={3705--3718},
  year={2017},
  publisher={IEEE}
}

@article{reis2026enforcing,
  title={Enforcing Current-State Opacity and Utility of Cyber-Physical Systems Using Multipath Event Routing},
  author={Reis, Lucas NR and Carvalho, Lilian K and Moreira, Marcos V},
  journal={IEEE Transactions on Automatic Control},
  year={2026},
  publisher={IEEE}
}

@article{ji2019opacity,
  title={Opacity enforcement using nondeterministic publicly known edit functions},
  author={Ji, Yiding and Yin, Xiang and Lafortune, St{\'e}phane},
  journal={IEEE Transactions on Automatic Control},
  volume={64},
  number={10},
  pages={4369--4376},
  year={2019},
  publisher={IEEE}
}

@article{tai2023privacy,
  title={Privacy-preserving co-synthesis against sensor--actuator eavesdropping intruder},
  author={Tai, Ruochen and Lin, Liyong and Zhu, Yuting and Su, Rong},
  journal={Automatica},
  volume={150},
  pages={110860},
  year={2023},
  publisher={Elsevier}
}

@article{duan2023edit,
  title={Edit mechanism synthesis for opacity enforcement under uncertain observations},
  author={Duan, Wei and Liu, Ruotian and Fanti, Maria Pia and Hadjicostis, Christoforos N and Li, Zhiwu},
  journal={IEEE Control Systems Letters},
  volume={7},
  pages={2041--2046},
  year={2023},
  publisher={IEEE}
}

@article{li2023opacity,
  title={Opacity enforcement in discrete event systems using extended insertion functions under inserted language constraints},
  author={Li, Xiaoyan and Hadjicostis, Christoforos N and Li, Zhiwu},
  journal={IEEE Transactions on Automatic Control},
  volume={68},
  number={11},
  pages={6797--6803},
  year={2023},
  publisher={IEEE}
}

@article{zheng2025enforcement,
  title={Enforcement of Opacity for Interval Weighted Discrete-Event System by Supervisory Control},
  author={Zheng, Yiwei and Lai, Aiwen and Lan, Weiyao and Yu, Xiao and Su, Rong},
  journal={IEEE Transactions on Automatic Control},
  year={2025},
  publisher={IEEE}
}

@article{liu2023opacity,
  title={Opacity enforcement via greedy privately-and-publicly known insertion functions},
  author={Liu, Rongjian and Lu, Jianquan and Liu, Yang and Yin, Xiang and Hadjicostis, Christoforos N},
  journal={IEEE Transactions on Automatic Control},
  volume={69},
  number={4},
  pages={2500--2506},
  year={2023},
  publisher={IEEE}
}

@article{duan2025obfuscation,
  title={Obfuscation mechanism for simultaneous public event information release and private event information hiding in discrete event systems},
  author={Duan, Wei and Hadjicostis, Christoforos N and Li, Zhiwu},
  journal={Information Sciences},
  volume={690},
  pages={121554},
  year={2025},
  publisher={Elsevier}
}

@article{keroglou2020embedded,
  title={Embedded insertion functions for opacity enforcement},
  author={Keroglou, Christoforos and Lafortune, St{\'e}phane},
  journal={IEEE Transactions on Automatic Control},
  volume={66},
  number={9},
  pages={4184--4191},
  year={2020},
  publisher={IEEE}
}

@inproceedings{wintenberg2022dynamic,
  title={A dynamic obfuscation framework for security and utility},
  author={Wintenberg, Andrew and Blischke, Matthew and Lafortune, St{\'e}phane and Ozay, Necmiye},
  booktitle={2022 ACM/IEEE 13th International Conference on Cyber-Physical Systems (ICCPS)},
  pages={236--246},
  year={2022},
  organization={IEEE}
}

@article{li2024opacity,
  title={Opacity Enforcement in Discrete Event Systems Using Modification Functions},
  author={Li, Xiaoyan and Hadjicostis, Christoforos N and Li, Zhiwu},
  journal={IEEE Transactions on Automation Science and Engineering},
  volume={22},
  pages={3252--3264},
  year={2024},
  publisher={IEEE}
}

@article{liu2022enforcement,
  title={Enforcement for infinite-step opacity and K-step opacity via insertion mechanism},
  author={Liu, Rongjian and Lu, Jianquan},
  journal={Automatica},
  volume={140},
  pages={110212},
  year={2022},
  publisher={Elsevier}
}

@article{li2021extended,
  title={Extended insertion functions for opacity enforcement in discrete-event systems},
  author={Li, Xiaoyan and Hadjicostis, Christoforos N and Li, Zhiwu},
  journal={IEEE Transactions on Automatic Control},
  volume={67},
  number={10},
  pages={5289--5303},
  year={2021},
  publisher={IEEE}
}

@article{wu2015synthesis,
  title={Synthesis of optimal insertion functions for opacity enforcement},
  author={Wu, Yi-Chin and Lafortune, St{\'e}phane},
  journal={IEEE Transactions on Automatic Control},
  volume={61},
  number={3},
  pages={571--584},
  year={2015},
  publisher={IEEE}
}

@article{ji2018enforcement,
  title={Enforcement of opacity by public and private insertion functions},
  author={Ji, Yiding and Wu, Yi-Chin and Lafortune, St{\'e}phane},
  journal={Automatica},
  volume={93},
  pages={369--378},
  year={2018},
  publisher={Elsevier}
}

@article{wu2014synthesis,
  title={Synthesis of insertion functions for enforcement of opacity security properties},
  author={Wu, Yi-Chin and Lafortune, St{\'e}phane},
  journal={Automatica},
  volume={50},
  number={5},
  pages={1336--1348},
  year={2014},
  publisher={Elsevier}
}

@article{wintenberg2025integrating,
  title={Integrating Obfuscation and Control for Privacy},
  author={Wintenberg, Andrew and Ozay, Necmiye and Lafortune, St{\'e}phane},
  journal={IEEE Transactions on Automatic Control},
  year={2025},
  publisher={IEEE}
}

@article{sears2018computing,
  title={Computing observers from observation policies in discrete-event systems},
  author={Sears, David and Rudie, Karen},
  journal={Discrete Event Dynamic Systems},
  volume={28},
  number={4},
  pages={509--537},
  year={2018},
  publisher={Springer}
}

@article{sears2016minimal,
  title={Minimal sensor activation and minimal communication in discrete-event systems},
  author={Sears, David and Rudie, Karen},
  journal={Discrete Event Dynamic Systems},
  volume={26},
  number={2},
  pages={295--349},
  year={2016},
  publisher={Springer}
}

@article{behinaein2019optimal,
  title={Optimal information release for mixed opacity in discrete-event systems},
  author={Behinaein, Behnam and Lin, Feng and Rudie, Karen},
  journal={IEEE Transactions on Automation Science and Engineering},
  volume={16},
  number={4},
  pages={1960--1970},
  year={2019},
  publisher={IEEE}
}

@article{yin2019general,
  title={A general approach for optimizing dynamic sensor activation for discrete event systems},
  author={Yin, Xiang and Lafortune, St{\'e}phane},
  journal={Automatica},
  volume={105},
  pages={376--383},
  year={2019},
  publisher={Elsevier}
}

@article{yin2017new,
  title={A new approach for the verification of infinite-step and K-step opacity using two-way observers},
  author={Yin, Xiang and Lafortune, St{\'e}phane},
  journal={Automatica},
  volume={80},
  pages={162--171},
  year={2017},
  publisher={Elsevier}
}

@article{yang2022secure,
  title={Secure your intention: On notions of pre-opacity in discrete-event systems},
  author={Yang, Shuo and Yin, Xiang},
  journal={IEEE Transactions on Automatic Control},
  volume={68},
  number={8},
  pages={4754--4766},
  year={2022},
  publisher={IEEE}
}

@article{miao2025always,
  title={Always guarding you: Strong initial-and-final-state opacity of discrete-event systems},
  author={Miao, Shaowen and Lai, Aiwen and Komenda, Jan},
  journal={Automatica},
  volume={173},
  pages={112085},
  year={2025},
  publisher={Elsevier}
}

@article{zhao2025no,
  title={No-regret path planning for temporal logic tasks in partially-known environments},
  author={Zhao, Jianing and Zhu, Keyi and Feng, Mingyang and Li, Shaoyuan and Yin, Xiang},
  journal={The International Journal of Robotics Research},
  volume={44},
  number={9},
  pages={1526--1552},
  year={2025},
  publisher={SAGE Publications Sage UK: London, England}
}

@article{han2023strong,
  title={Strong current-state and initial-state opacity of discrete-event systems},
  author={Han, Xiaoguang and Zhang, Kuize and Zhang, Jiahui and Li, Zhiwu and Chen, Zengqiang},
  journal={Automatica},
  volume={148},
  pages={110756},
  year={2023},
  publisher={Elsevier}
}

@article{saboori2011verification,
  title={Verification of infinite-step opacity and complexity considerations},
  author={Saboori, Anooshiravan and Hadjicostis, Christoforos N},
  journal={IEEE Transactions on Automatic Control},
  volume={57},
  number={5},
  pages={1265--1269},
  year={2011},
  publisher={IEEE}
}

@article{saboori2011verification2,
  title={Verification of $ K $-step opacity and analysis of its complexity},
  author={Saboori, Anooshiravan and Hadjicostis, Christoforos N},
  journal={IEEE Transactions on Automation Science and Engineering},
  volume={8},
  number={3},
  pages={549--559},
  year={2011},
  publisher={IEEE}
}

@article{balun2021comparing, 
  title={Comparing the notions of opacity for discrete-event systems},
  author={Balun, Ji{\v{r}}{\'\i} and Masopust, Tom{\'a}{\v{s}}},
  journal={Discrete Event Dynamic Systems},
  volume={31},
  number={4},
  pages={553--582},
  year={2021},
  publisher={Springer}
}

@article{wu2013comparative,
  title={Comparative analysis of related notions of opacity in centralized and coordinated architectures},
  author={Wu, Yi-Chin and Lafortune, St{\'e}phane},
  journal={Discrete Event Dynamic Systems},
  volume={23},
  number={3},
  pages={307--339},
  year={2013},
  publisher={Springer}
}

@article{masopust2019complexity,
  title={Complexity of detectability, opacity and A-diagnosability for modular discrete event systems},
  author={Masopust, Tom{\'a}{\v{s}} and Yin, Xiang},
  journal={Automatica},
  volume={101},
  pages={290--295},
  year={2019},
  publisher={Elsevier}
}

@article{yin2017verification,
  title={Verification complexity of a class of observational properties for modular discrete events systems},
  author={Yin, Xiang and Lafortune, St{\'e}phane},
  journal={Automatica},
  volume={83},
  pages={199--205},
  year={2017},
  publisher={Elsevier}
}

@article{mohajerani2019transforming,
  title={Transforming opacity verification to nonblocking verification in modular systems},
  author={Mohajerani, Sahar and Lafortune, St{\'e}phane},
  journal={IEEE Transactions on Automatic Control},
  volume={65},
  number={4},
  pages={1739--1746},
  year={2019},
  publisher={IEEE}
}

@article{lennartson2022state,
  title={State-labeled safety analysis of modular observers for opacity verification},
  author={Lennartson, Bengt and Noori-Hosseini, Mona and Hadjicostis, Christoforos N},
  journal={IEEE Control Systems Letters},
  volume={6},
  pages={2936--2941},
  year={2022},
  publisher={IEEE}
}

@article{kalat2021modular,
  title={Modular verification of opacity for interconnected control systems via barrier certificates},
  author={Kalat, Shadi Tasdighi and Liu, Siyuan and Zamani, Majid},
  journal={IEEE Control Systems Letters},
  volume={6},
  pages={890--895},
  year={2021},
  publisher={IEEE}
}

@inproceedings{tong2019current,
  title={Current-state opacity verification in modular discrete event systems},
  author={Tong, Yin and Lan, Hao},
  booktitle={2019 IEEE 58th Conference on Decision and Control (CDC)},
  pages={7665--7670},
  year={2019},
  organization={IEEE}
}

@article{wang2010minimization,
  title={Minimization of dynamic sensor activation in discrete event systems for the purpose of control},
  author={Wang, Weilin and Lafortune, St{\'e}phane and Lin, Feng and Girard, Anouck R},
  journal={IEEE Transactions on Automatic Control},
  volume={55},
  number={11},
  pages={2447--2461},
  year={2010},
  publisher={IEEE}
}

@inproceedings{udupa2025synthesis,
  title={Synthesis of Dynamic Masks for Information-Theoretic Opacity in Stochastic Systems},
  author={Udupa, Sumukha and Shi, Chongyang and Fu, Jie},
  booktitle={Proceedings of the ACM/IEEE 16th International Conference on Cyber-Physical Systems (with CPS-IoT Week 2025)},
  pages={1--12},
  year={2025}
}

@article{berard2018opacity,
  title={Opacity for linear constraint Markov chains},
  author={B{\'e}rard, B{\'e}atrice and Kouchnarenko, Olga and Mullins, John and Sassolas, Mathieu},
  journal={Discrete Event Dynamic Systems},
  volume={28},
  number={1},
  pages={83--108},
  year={2018},
  publisher={Springer}
}

@inproceedings{cui2023security,
  title={Security-aware reinforcement learning under linear temporal logic specifications},
  author={Cui, Bohan and Zhu, Keyi and Li, Shaoyuan and Yin, Xiang},
  booktitle={2023 IEEE International Conference on Robotics and Automation (ICRA)},
  pages={12367--12373},
  year={2023},
  organization={IEEE}
}

@article{mao2024minimal,
  title={Minimal Sensor Activation for Detectable Networked Discrete Event Systems},
  author={Mao, Anjie and Zhang, Yuchen and Wang, Zheming and Chen, Bo},
  journal={IEEE Control Systems Letters},
  volume={8},
  pages={340--345},
  year={2024},
  publisher={IEEE}
}

@article{yin2016synthesis,
  title={Synthesis of maximally-permissive supervisors for the range control problem},
  author={Yin, Xiang and Lafortune, St{\'e}phane},
  journal={IEEE Transactions on Automatic Control},
  volume={62},
  number={8},
  pages={3914--3929},
  year={2016},
  publisher={IEEE}
}

@article{chen2023you,
  title={You don't know when i will arrive: Unpredictable controller synthesis for temporal logic tasks},
  author={Chen, Yu and Yang, Shuo and Mangharam, Rahul and Yin, Xiang},
  journal={IFAC-PapersOnLine},
  volume={56},
  number={2},
  pages={3591--3597},
  year={2023},
  publisher={Elsevier}
}

@article{tai2025privacy,
  title={Privacy-Preserving Supervisory Control for Data Opacity Enforcement},
  author={Tai, Ruochen and Lin, Liyong and Su, Rong and Ge, Shuzhi Sam},
  journal={IEEE Transactions on Industrial Informatics},
  year={2025},
  publisher={IEEE}
}

@article{berard2015quantifying,
  title={Quantifying opacity},
  author={B{\'e}rard, B{\'e}atrice and Mullins, John and Sassolas, Mathieu},
  journal={Mathematical Structures in Computer Science},
  volume={25},
  number={2},
  pages={361--403},
  year={2015},
  publisher={Cambridge University Press}
}

@article{chen2016quantification,
  title={Quantification of secrecy in partially observed stochastic discrete event systems},
  author={Chen, Jun and Ibrahim, Mariam and Kumar, Ratnesh},
  journal={IEEE Transactions on Automation Science and Engineering},
  volume={14},
  number={1},
  pages={185--195},
  year={2016},
  publisher={IEEE}
}

@inproceedings{yin2019opacity,
  title={Opacity of networked supervisory control systems over insecure multiple channel networks},
  author={Yin, Xiang and Li, Shaoyuan},
  booktitle={2019 IEEE 58th Conference on Decision and Control (CDC)},
  pages={7641--7646},
  year={2019},
  organization={IEEE}
}

@article{xie2021opacity,
  title={Opacity enforcing supervisory control using nondeterministic supervisors},
  author={Xie, Yifan and Yin, Xiang and Li, Shaoyuan},
  journal={IEEE Transactions on Automatic Control},
  volume={67},
  number={12},
  pages={6567--6582},
  year={2021},
  publisher={IEEE}
}

@article{cui2025prediction,
  title={On Prediction-Based Properties of Discrete-Event Systems: Notions, Applications and Supervisor Synthesis},
  author={Cui, Bohan and Chen, Yu and Giua, Alessandro and Yin, Xiang},
  journal={arXiv preprint arXiv:2510.04616},
  year={2025}
}

@article{xie2023optimal,
  title={Optimal synthesis of opacity-enforcing supervisors for qualitative and quantitative specifications},
  author={Xie, Yifan and Li, Shaoyuan and Yin, Xiang},
  journal={IEEE Transactions on Automatic Control},
  volume={69},
  number={8},
  pages={4958--4973},
  year={2023},
  publisher={IEEE}
}

@inproceedings{xie2020note,
  title={Supervisory Control of Discrete-Event Systems for Infinite-Step Opacity},
  author={Xie, Yifan and Yin, Xiang},
  booktitle={2020 American Control Conference (ACC)},
  pages={3665--3671},
  year={2020},
  organization={IEEE}
}

@inproceedings{xie2021secure,
  title={Secure-by-construction controller synthesis for stochastic systems under linear temporal logic specifications},
  author={Xie, Yifan and Yin, Xiang and Li, Shaoyuan and Zamani, Majid},
  booktitle={2021 60th IEEE Conference on Decision and Control (CDC)},
  pages={7015--7021},
  year={2021},
  organization={IEEE}
}

@article{dai2026optimal,
  title={Optimal control strategy for k-level opaque systems based on the optimization method of multi-selection problems at each stage},
  author={Dai, Yinyin and Wang, Fei and Luo, Jiliang},
  journal={Discrete Event Dynamic Systems},
  volume={36},
  number={1},
  pages={8},
  year={2026},
  publisher={Springer}
}

@article{moulton2022using,
  title={Using subobservers to synthesize opacity-enforcing supervisors},
  author={Moulton, Richard Hugh and Hamgini, Behnam Behinaein and Khouzani, Zahra Abedi and Meira-G{\'o}es, R{\^o}mulo and Wang, Fei and Rudie, Karen},
  journal={Discrete Event Dynamic Systems},
  volume={32},
  number={4},
  pages={611--640},
  year={2022},
  publisher={Springer}
}

@article{tong2018current,
  title={Current-state opacity enforcement in discrete event systems under incomparable observations},
  author={Tong, Yin and Li, Zhiwu and Seatzu, Carla and Giua, Alessandro},
  journal={Discrete Event Dynamic Systems},
  volume={28},
  number={2},
  pages={161--182},
  year={2018},
  publisher={Springer}
}

@inproceedings{yin2015new,
  title={A new approach for synthesizing opacity-enforcing supervisors for partially-observed discrete-event systems},
  author={Yin, Xiang and Lafortune, St{\'e}phane},
  booktitle={2015 American Control Conference (ACC)},
  pages={377--383},
  year={2015},
  organization={IEEE}
}

@article{yin2015uniform,
  title={A uniform approach for synthesizing property-enforcing supervisors for partially-observed discrete-event systems},
  author={Yin, Xiang and Lafortune, St{\'e}phane},
  journal={IEEE Transactions on Automatic Control},
  volume={61},
  number={8},
  pages={2140--2154},
  year={2015},
  publisher={IEEE}
}

@article{darondeau2015enforcing,
  title={Enforcing opacity of regular predicates on modal transition systems},
  author={Darondeau, Philippe and Marchand, Herv{\'e} and Ricker, Laurie},
  journal={Discrete Event Dynamic Systems},
  volume={25},
  number={1},
  pages={251--270},
  year={2015},
  publisher={Springer}
}

@article{yang2021opacity,
  title={Opacity of networked supervisory control systems over insecure communication channels},
  author={Yang, Shuo and Hou, Junyao and Yin, Xiang and Li, Shaoyuan},
  journal={IEEE Transactions on Control of Network Systems},
  volume={8},
  number={2},
  pages={884--896},
  year={2021},
  publisher={IEEE}
}

@article{takai2008formula,
  title={A formula for the supremal controllable and opaque sublanguage arising in supervisory control},
  author={Takai, Shigemasa and Oka, Yusuke},
  journal={SICE Journal of Control, Measurement, and System Integration},
  volume={1},
  number={4},
  pages={307--311},
  year={2008},
  publisher={Taylor \& Francis}
}

@article{saboori2011opacity,
  title={Opacity-enforcing supervisory strategies via state estimator constructions},
  author={Saboori, Anooshiravan and Hadjicostis, Christoforos N},
  journal={IEEE Transactions on Automatic Control},
  volume={57},
  number={5},
  pages={1155--1165},
  year={2011},
  publisher={IEEE}
}

@article{liu2022secure,
  title={Secure-by-construction synthesis of cyber-physical systems},
  author={Liu, Siyuan and Trivedi, Ashutosh and Yin, Xiang and Zamani, Majid},
  journal={Annual Reviews in Control},
  volume={53},
  pages={30--50},
  year={2022},
  publisher={Elsevier}
}

@article{guo2020overview, 
  title={Overview of opacity in discrete event systems},
  author={Guo, Ye and Jiang, Xiaoning and Guo, Chen and Wang, Shouguang and Karoui, Oussama},
  journal={IEEE Access},
  volume={8},
  pages={48731--48741},
  year={2020},
  publisher={IEEE}
}

@article{oliveira2023classification,
  title={A classification of cybersecurity strategies in the context of discrete event systems},
  author={Oliveira, Samuel and Leal, Andre B and Teixeira, Marcelo and Lopes, Yuri K},
  journal={Annual reviews in Control},
  volume={56},
  pages={100907},
  year={2023},
  publisher={Elsevier}
}

@article{lafortune2018history,
  title={On the history of diagnosability and opacity in discrete event systems},
  author={Lafortune, St{\'e}phane and Lin, Feng and Hadjicostis, Christoforos N},
  journal={Annual Reviews in Control},
  volume={45},
  pages={257--266},
  year={2018},
  publisher={Elsevier}
}

@article{dubreil2010supervisory,
  title={Supervisory control for opacity},
  author={Dubreil, J{\'e}r{\'e}my and Darondeau, Philippe and Marchand, Herv{\'e}},
  journal={IEEE Transactions on Automatic Control},
  volume={55},
  number={5},
  pages={1089--1100},
  year={2010},
  publisher={IEEE}
}

@article{wonham2018supervisory,
  title={Supervisory control of discrete-event systems: A brief history},
  author={Wonham, Walter Murry and Cai, Kai and Rudie, Karen},
  journal={Annual Reviews in Control},
  volume={45},
  pages={250--256},
  year={2018},
  publisher={Elsevier}
}

@article{wintenberg2022general,
  title={A general language-based framework for specifying and verifying notions of opacity},
  author={Wintenberg, Andrew and Blischke, Matthew and Lafortune, St{\'e}phane and Ozay, Necmiye},
  journal={Discrete Event Dynamic Systems},
  volume={32},
  number={2},
  pages={253--289},
  year={2022},
  publisher={Springer}
}

@article{yang2022current,
  title={Current-state opacity and initial-state opacity of modular discrete event systems},
  author={Yang, Jingkai and Deng, Weilin and Qiu, Daowen},
  journal={International Journal of Control},
  volume={95},
  number={11},
  pages={3037--3049},
  year={2022},
  publisher={Taylor \& Francis}
}

@article{ramadge1987supervisory,
  title={Supervisory control of a class of discrete event processes},
  author={Ramadge, Peter J and Wonham, W Murray},
  journal={SIAM journal on control and optimization},
  volume={25},
  number={1},
  pages={206--230},
  year={1987},
  publisher={SIAM}
}

@inproceedings{wu2018parameter,
  title={Parameter and insertion function co-synthesis for opacity enhancement in parametric stochastic discrete event systems},
  author={Wu, Bo and Liu, Zhiyu and Lin, Hai},
  booktitle={2018 Annual American Control Conference (ACC)},
  pages={3032--3037},
  year={2018},
  organization={IEEE}
}

@inproceedings{andre2024bright,
  title={The bright side of timed opacity},
  author={Andr{\'e}, {\'E}tienne and D{\'e}pernet, Sarah and Lefaucheux, Engel},
  booktitle={International Conference on Formal Engineering Methods},
  pages={51--69},
  year={2024},
  organization={Springer}
}

@article{andre2022guaranteeing,
  title={Guaranteeing timed opacity using parametric timed model checking},
  author={Andr{\'e}, {\'E}tienne and Lime, Didier and Marinho, Dylan and Sun, Jun},
  journal={ACM Transactions on Software Engineering and Methodology (TOSEM)},
  volume={31},
  number={4},
  pages={1--36},
  year={2022},
  publisher={ACM New York, NY}
}

@inproceedings{cassez2009dark,
  title={The dark side of timed opacity},
  author={Cassez, Franck},
  booktitle={International conference on information security and assurance},
  pages={21--30},
  year={2009},
  organization={Springer}
}

@article{dong2024verification,
  title={On the verification of detectability for timed discrete event systems},
  author={Dong, Weijie and Zhang, Kuize and Li, Shaoyuan and Yin, Xiang},
  journal={Automatica},
  volume={164},
  pages={111644},
  year={2024},
  publisher={Elsevier}
}

@article{deng2025opacity,
  title={Opacity of Extended Finite Automata With Event Parameters},
  author={Deng, Weilin and Qiu, Daowen and Yang, Jingkai},
  journal={IEEE Transactions on Automatic Control},
  year={2025},
  publisher={IEEE}
}

@article{li2024verification,
  title={Verification of state-based timed opacity for constant-time labeled automata},
  author={Li, Jun and Lefebvre, Dimitri and Hadjicostis, Christoforos N and Li, Zhiwu},
  journal={IEEE Transactions on Automatic Control},
  year={2024},
  publisher={IEEE}
}

@article{zhang2024state,
  title={State-based opacity of labeled real-time automata},
  author={Zhang, Kuize},
  journal={Theoretical Computer Science},
  volume={987},
  pages={114373},
  year={2024},
  publisher={Elsevier}
}

@article{wang2018opacity,
  title={The opacity of real-time automata},
  author={Wang, Lingtai and Zhan, Naijun and An, Jie},
  journal={IEEE Transactions on Computer-Aided Design of Integrated Circuits and Systems},
  volume={37},
  number={11},
  pages={2845--2856},
  year={2018},
  publisher={IEEE}
}

@article{lai2021initial,
  title={Initial-state detectability and initial-state opacity of unambiguous weighted automata},
  author={Lai, Aiwen and Lahaye, S{\'e}bastien and Li, Zhiwu},
  journal={Automatica},
  volume={127},
  pages={109490},
  year={2021},
  publisher={Elsevier}
}

@article{deng2025initial,
  title={Initial-Location Opacity and Infinite-Step Opacity of Timed Automata With Integer Resets},
  author={Deng, Weilin and Qiu, Daowen and Yang, Jingkai},
  journal={IEEE Control Systems Letters},
  year={2025},
  publisher={IEEE}
}

@article{ammar2021bounded,
  title={Bounded opacity for timed systems},
  author={Ammar, Ikhlass and El Touati, Yamen and Yeddes, Moez and Mullins, John},
  journal={Journal of Information Security and Applications},
  volume={61},
  pages={102926},
  year={2021},
  publisher={Elsevier}
}

@article{tong2016verification,
  title={Verification of state-based opacity using Petri nets},
  author={Tong, Yin and Li, Zhiwu and Seatzu, Carla and Giua, Alessandro},
  journal={IEEE Transactions on Automatic Control},
  volume={62},
  number={6},
  pages={2823--2837},
  year={2016},
  publisher={IEEE}
}

@article{bryans2005modelling,
  title={Modelling opacity using Petri nets},
  author={Bryans, Jeremy W and Koutny, Maciej and Ryan, Peter YA},
  journal={Electronic Notes in Theoretical Computer Science},
  volume={121},
  pages={101--115},
  year={2005},
  publisher={Elsevier}
}

@article{berard2018complexity,
  title={The complexity of diagnosability and opacity verification for Petri nets},
  author={B{\'e}rard, B{\'e}atrice and Haar, Stefan and Schmitz, Sylvain and Schwoon, Stefan},
  journal={Fundamenta Informaticae},
  volume={161},
  number={4},
  pages={317--349},
  year={2018},
  publisher={SAGE Publications Sage UK: London, England}
}

@article{tong2017decidability,
  title={Decidability of opacity verification problems in labeled Petri net systems},
  author={Tong, Yin and Li, Zhiwu and Seatzu, Carla and Giua, Alessandro},
  journal={Automatica},
  volume={80},
  pages={48--53},
  year={2017},
  publisher={Elsevier}
}

@article{cong2018line,
  title={On-line verification of current-state opacity by Petri nets and integer linear programming},
  author={Cong, Xuya and Fanti, Maria Pia and Mangini, Agostino Marcello and Li, Zhiwu},
  journal={Automatica},
  volume={94},
  pages={205--213},
  year={2018},
  publisher={Elsevier}
}

@article{dong2025k,
  title={$ K $-step Opacity Verification and Enforcement of Time Labeled Petri Net Systems},
  author={Dong, Yifan and Lefebvre, Dimitri and Li, Zhiwu},
  journal={IEEE Transactions on Automatic Control},
  year={2025},
  publisher={IEEE}
}

@article{kohler2024enforcement,
  title={Enforcement of Current-State Opacity in Signal Interpreted Petri Nets},
  author={K{\"o}hler, Andreas and Marijan, Pascal and Zhang, Ping},
  journal={IEEE Transactions on Automatic Control},
  volume={69},
  number={11},
  pages={8104--8111},
  year={2024},
  publisher={IEEE}
}

@article{zhu2022online,
  title={Online verification of K-step opacity by Petri nets in centralized and decentralized structures},
  author={Zhu, Guanghui and Li, Zhiwu and Wu, Naiqi},
  journal={Automatica},
  volume={145},
  pages={110528},
  year={2022},
  publisher={Elsevier}
}

@article{tong2022verification,
  title={Verification of K-step and infinite-step opacity of bounded labeled Petri nets},
  author={Tong, Yin and Lan, Hao and Seatzu, Carla},
  journal={Automatica},
  volume={140},
  pages={110221},
  year={2022},
  publisher={Elsevier}
}

@article{basile2023assessment,
  title={Assessment of initial-state-opacity in live and bounded labeled Petri net systems via optimization techniques},
  author={Basile, Francesco and De Tommasi, Gianmaria and Motta, Carlo},
  journal={Automatica},
  volume={152},
  pages={110911},
  year={2023},
  publisher={Elsevier}
}

@article{ma2020marking,
  title={Marking predictability and prediction in labeled Petri nets},
  author={Ma, Ziyue and Yin, Xiang and Li, Zhiwu},
  journal={IEEE Transactions on Automatic Control},
  volume={66},
  number={8},
  pages={3608--3623},
  year={2020},
  publisher={IEEE}
}

@article{basile2022necessary,
  title={Necessary and sufficient condition to assess initial-state-opacity in live bounded and reversible discrete event systems},
  author={Basile, Francesco and De Tommasi, Gianmaria and Motta, Carlo and Sterle, Claudio},
  journal={IEEE Control Systems Letters},
  volume={6},
  pages={2683--2688},
  year={2022},
  publisher={IEEE}
}

@article{masopust2019deciding,
  title={Deciding detectability for labeled Petri nets},
  author={Masopust, Tom{\'a}{\v{s}} and Yin, Xiang},
  journal={Automatica},
  volume={104},
  pages={238--241},
  year={2019},
  publisher={Elsevier}
}

@article{ma2021marking,
  title={Marking diagnosability verification in labeled Petri nets},
  author={Ma, Ziyue and Yin, Xiang and Li, Zhiwu},
  journal={Automatica},
  volume={131},
  pages={109713},
  year={2021},
  publisher={Elsevier}
}

@article{yin2017decidability,
  title={On the decidability and complexity of diagnosability for labeled Petri nets},
  author={Yin, Xiang and Lafortune, St{\'e}phane},
  journal={IEEE Transactions on Automatic Control},
  volume={62},
  number={11},
  pages={5931--5938},
  year={2017},
  publisher={IEEE}
}

@article{ma2016basis,
  title={Basis marking representation of Petri net reachability spaces and its application to the reachability problem},
  author={Ma, Ziyue and Tong, Yin and Li, Zhiwu and Giua, Alessandro},
  journal={IEEE Transactions on Automatic Control},
  volume={62},
  number={3},
  pages={1078--1093},
  year={2016},
  publisher={IEEE}
}

@article{yin2020approximate,
  title={On approximate opacity of cyber-physical systems},
  author={Yin, Xiang and Zamani, Majid and Liu, Siyuan},
  journal={IEEE Transactions on Automatic Control},
  volume={66},
  number={4},
  pages={1630--1645},
  year={2020},
  publisher={IEEE}
}

@article{ramasubramanian2019notions,
  title={Notions of centralized and decentralized opacity in linear systems},
  author={Ramasubramanian, Bhaskar and Cleaveland, Rance and Marcus, Steven I},
  journal={IEEE Transactions on Automatic Control},
  volume={65},
  number={4},
  pages={1442--1455},
  year={2019},
  publisher={IEEE}
}

@article{an2019opacity,
  title={Opacity enforcement for confidential robust control in linear cyber-physical systems},
  author={An, Liwei and Yang, Guang-Hong},
  journal={IEEE Transactions on Automatic Control},
  volume={65},
  number={3},
  pages={1234--1241},
  year={2019},
  publisher={IEEE}
}

@article{zhang2019opacity,
  title={Opacity of nondeterministic transition systems: A (bi) simulation relation approach},
  author={Zhang, Kuize and Yin, Xiang and Zamani, Majid},
  journal={IEEE Transactions on Automatic Control},
  volume={64},
  number={12},
  pages={5116--5123},
  year={2019},
  publisher={IEEE}
}

@article{john2024opacity,
  title={Opacity vs. Security in Linear Dynamical Systems},
  author={John, Varkey M and Katewa, Vaibhav},
  journal={IEEE Transactions on Automatic Control},
  year={2024},
  publisher={IEEE}
}

@article{hou2022abstraction,
  title={Abstraction-based verification of approximate preopacity for control systems},
  author={Hou, Junyao and Liu, Siyuan and Yin, Xiang and Zamani, Majid},
  journal={IEEE Control Systems Letters},
  volume={7},
  pages={1087--1092},
  year={2022},
  publisher={IEEE}
}

@inproceedings{hou2023abstraction,
  title={Abstraction-Based Synthesis of Controllers for Approximate Opacity},
  author={Hou, Junyao and Liu, Siyuan and Yin, Xiang and Zamani, Majid},
  booktitle={2023 62nd IEEE Conference on Decision and Control (CDC)},
  pages={7930--7936},
  year={2023},
  organization={IEEE}
}

@article{qiao2025approximate,
  title={Approximate Opacity and Its Enforcement for Cyber-Physical Systems Based on Reinforcement Q-Learning},
  author={Qiao, Xinxin and Qi, Yiwen and Zheng, Zonghua and Wang, Wu},
  journal={IEEE Transactions on Industrial Cyber-Physical Systems},
  volume={4},
  pages={1--11},
  year={2025},
  publisher={IEEE}
}

@inproceedings{hou2019abstraction,
  title={Abstraction-based synthesis of opacity-enforcing controllers using alternating simulation relations},
  author={Hou, Junyao and Yin, Xiang and Li, Shaoyuan and Zamani, Majid},
  booktitle={2019 IEEE 58th Conference on Decision and Control (CDC)},
  pages={7653--7658},
  year={2019},
  organization={IEEE}
}

@article{zhong2025secure,
  title={Secure-by-construction synthesis for control systems},
  author={Zhong, Bingzhuo and Liu, Siyuan and Caccamo, Marco and Zamani, Majid},
  journal={IEEE Transactions on Automatic Control},
  year={2025},
  publisher={IEEE}
}

@article{mizoguchi2021abstraction,
  title={Abstraction-based control under quantized observation with approximate opacity using symbolic control barrier functions},
  author={Mizoguchi, Masashi and Ushio, Toshimitsu},
  journal={IEEE Control Systems Letters},
  volume={6},
  pages={2222--2227},
  year={2021},
  publisher={IEEE}
}

@article{liu2024approximate,
  title={On approximate opacity of stochastic control systems},
  author={Liu, Siyuan and Yin, Xiang and Dimarogonas, Dimos V and Zamani, Majid},
  journal={IEEE Transactions on Automatic Control},
  year={2024},
  publisher={IEEE}
}

@article{liu2021compositional,
  title={Compositional synthesis of opacity-preserving finite abstractions for interconnected systems},
  author={Liu, Siyuan and Zamani, Majid},
  journal={Automatica},
  volume={131},
  pages={109745},
  year={2021},
  publisher={Elsevier}
}

\appendix

\end{document}